\documentclass[oneside,a4paper,11pt,shownumbers]{article}
\usepackage{latexsym}
\usepackage{euscript}
\usepackage{epsfig,amsmath,amssymb}

\def\be{\begin{equation}}
\def\ee{\end{equation}}
\def\bea{\begin{eqnarray}}
\def\eea{\end{eqnarray}}

\long\def\symbolfootnote[#1]#2{\begingroup%
\def\thefootnote{\fnsymbol{footnote}}\footnote[#1]{#2}\endgroup} 

 \large\normalsize

\begin{document}

\begin{center}

{\Large \bf Interacting $Q$-balls}

\vspace*{7mm} {Yves Brihaye $^{a}$
\symbolfootnote[1]{E-mail: yves.brihaye@umh.ac.be}
and  Betti Hartmann $^{b}$
\symbolfootnote[2]{E-mail:b.hartmann@jacobs-university.de}}
\vspace*{.25cm}

${}^{a)}${\it Facult\'e des Sciences, Universit\'e de Mons-Hainaut, 7000 Mons, Belgium}\\
${}^{b)}${\it School of Engineering and Science, Jacobs University Bremen, 28759 Bremen, Germany}\\

\vspace*{.3cm}
\end{center}

\begin{abstract}
We study non-topological solitons, so called $Q$-balls, which carry
a non-vanishing Noether charge and arise as lump solutions of
self-interacting complex scalar field models.
Explicit examples of new axially symmetric
non-spinning $Q$-ball solutions  that have not been studied so far are constructed numerically. 
These solutions can be interpreted as angular excitations of the fundamental $Q$-balls and are related to the 
spherical harmonics. Correspondingly, they have higher energy and their energy densities possess two local maxima 
on the positive $z$-axis. 

We also study two $Q$-balls interacting via a potential term in $3+1$ dimensions
and construct examples of stationary, solitonic-like objects in
$(3+1)$-dimensional flat space-time that consist of two interacting global scalar fields.
We concentrate on configurations composed of one spinning and one non-spinning $Q$-ball and study the parameter-dependence
of the energy and charges of the configuration.

In addition, we present numerical evidence that for fixed values of the coupling constants
two different types of 2-$Q$-ball solutions exist: solutions with defined parity, but also solutions
which are asymmetric with respect to reflexion through the $x$-$y$-plane.
\end{abstract}

\section{Introduction}
Solitons play an important role in many areas of physics. As classical solutions of non-linear field theories, they
are localised structures with finite energy, which are globally regular.
In general, one can distinguish topological and non-topological solitons.
While topological solitons \cite{ms} possess a conserved quantity, the topological charge, that stems (in most
cases) from the spontaneous symmetry breaking of the theory, non-topological solitons \cite{fls,lp} have a conserved Noether
charge that results from a symmetry of the Lagrangian. The standard example of  non-topological solitons
are $Q$-balls \cite{coleman}, which are solutions of theories with self-interacting complex scalar fields. These objects are stationary with an explicitely
time-dependent phase. The conserved Noether charge $Q$
is then related to the global phase invariance of the theory and is directly proportional
to the frequency. $Q$ can e.g. be interpreted as particle number \cite{fls}. 

While in standard scalar
field theories, it was shown
that a non-renormalisable $\Phi^6$-potential is necessary \cite{vw}, supersymmetric extensions of the
Standard Model (SM)  also possess $Q$-ball solutions \cite{kusenko}. In the latter case, several scalar fields
interact via complicated potentials. It was shown that cubic interaction terms that result from
Yukawa couplings in the superpotential and supersymmetry breaking terms lead to the existence of $Q$-balls
with non-vanishing baryon or lepton number or electric charge. These supersymmetric
$Q$-balls have been considered recently as possible candidates for baryonic dark matter 
\cite{dm} and their astrophysical implications have been discussed \cite{implications}.

Two interacting scalar fields are also interesting from another point of view.
Up until now, the number of explicit examples of stationary solitonic-like solutions that involve two interacting
global
scalar fields is small. An important example are superconducting strings, which are axially
symmetric in $2+1$ dimensions extended trivially into the $z$-direction \cite{witten}.
Axially symmetric generalisations in $3+1$ dimensions, so-called vortons, have been constructed in \cite{ls}. 
Note that all these solutions
have been constructed in models which have a renormalisable $\Phi^4$-potential.

Here, we study two interacting scalar fields in $3+1$ dimensions and construct explicit
examples of stationary solitonic-like axially symmetric solutions consisting of two global scalar fields.
While vortons possess one scalar field with an unbroken U(1)
symmetry
(the condensate field) and a scalar field whose U(1) is spontaneously
broken (the string field), we here consider two scalar fields with
unbroken U(1) symmetries.
One can thus see our model as the limit of vanishing
vacuum expectation value for the second scalar field. Then, stationary solitonic
like objects can be constructed explicitely. Note that the model in \cite{ls}
contains a renormalisable $\Phi^4$-potential, while we need a non-renormalisable $\Phi^6$-potential here. 
However, as stated in \cite{ls}, the explicit construction of vortons was done
using also a non-renormalisable potential which contains an interaction term of
the form $\Phi_1^6 \Phi_2^2$.

$Q$-ball solutions in $3+1$ dimensions have been first studied in detail in 
\cite{vw}. It was realised
that next to non-spinning $Q$-balls, which are spherically symmetric, spinning solutions
exist. These are axially symmetric with energy density of toroidal shape
and angular momentum $J=kQ$, where $Q$ is the Noether charge of the solution
and $k\in \mathbb{Z}$ corresponds to the winding around the $z$-axis. 
Approximated  solutions of the non-linear partial differential equations
were constructed in \cite{vw} by means of a truncated series in the spherical harmonics to describe
the angular part of the solutions. 
The full  partial differential equation was solved numerically in \cite{kk}. 
It was also realised in \cite{vw} that in each $k$-sector, parity-even ($P=+1$)
and parity-odd ($P=-1$) solutions exist. Parity-even and parity-odd 
refers to the fact that
the solution is symmetric and anti-symmetric, respectively with respect
to a reflection through the $x$-$y$-plane, i.e. under $\theta\rightarrow \pi-\theta$.

These two types of solutions are
closely related to the fact that the angular part of the solutions
constructed in \cite{vw,kk}
is connected to the spherical harmonic $Y_0^0(\theta,\varphi)$ for the spherically symmetric $Q$-ball,
to the spherical harmonic $Y_1^{1}(\theta,\varphi)$ for the spinning parity even ($P=+1$) solution
and to the spherical harmonic $Y_2^{1}(\theta,\varphi)$  for the parity  odd ($P=-1$) solution, respectively. 
Radially excited solutions of the spherically symmetric, non-spinning solution were also obtained.
These solutions are still spherically symmetric but the scalar field develops one or several nodes for $r\in ]0,\infty[$. 
In relation to the apparent connection of the angular part of the known solutions to the spherical harmonics,
it is  natural to investigate whether ``$\theta$-angular excitations'' of the $Q$-balls exist
in correspondence to the whole family of spherical harmonics  $Y_L^k(\theta,\varphi)$, $-L \leq k \leq L$.
This can further be motivated by the fact that,
in the small field limit where a linear approximation can be used,
the field equation describing the $Q$-ball
becomes a standard harmonic equation that can be
solved by separation of variables and whose fundamental 
solutions are given in terms of spherical harmonics for the angular part.
Of course, it has to be checked whether this correspondence, 
expected from the linear limit,
still holds for the full, i.e. non-linear equation.

In the present paper, we present strong numerical arguments that new angularly 
excited solutions of the non-linear field equations exist and that the correspondence
between angular excitations of the $Q$-balls and spherical harmonics indeed holds. In addition to the solutions
corresponding to $Y_k^{ k}$ and $Y_k^{k-1}$ for $k=1,2,3$ presented in \cite{vw}
we have constructed solutions with angular dependence and symmetries corresponding to 
the spherical harmonics $Y_1^0$ and $Y_2^0$.  These solutions are non-spinning 
but constitute axially symmetric excitations with respect to the angular coordinate $\theta$. As expected, these new solutions
have higher energies and charges than the spherically symmetric solutions and we
would thus expect them to be unstable.
These solutions thus complete the already known spectrum of $Q$-ball solutions
and show that not only radial excitations of fundamental soliton solutions, but also angular excitations exist.

We also study two interacting $Q$-balls and put the emphasis on the interaction
between a non-spinning and a spinning $Q$-ball. In particular, we investigate the dependence
of the energy and the charges of the solution on the interaction parameter and the frequencies, respectively.

Next to parity-even and parity-odd solutions, we also construct solutions that have no
defined parity with respect to reflexion through the $x$-$y$-plane.

The explicit construction of solutions with two interacting complex scalar fields is surely of interest for the astrophysical implications of such objects, especially
for the construction of such objects in supersymmetric theories. Moreover, it 
adds to the spectrum of soliton solutions that e.g. possess
no definite parity.

The differential equations describing both excited as well as interacting
$Q$-balls are non-linear partial differential equations, which -to our knowledge- cannot
be solved analytically. We thus solve these equations numerically using an appropriate
PDE solver \cite{fidi}.

Our paper is organised as follows: in Section 2, we discuss the model
and give the equations and boundary conditions. In Section 3, we discuss
the new $Q$-ball solutions for $k=0$, while in Section 4, we present our results
for two interacting $Q$-balls. Section 5 contains our conclusions.

\section{The model}
In the following, we study a scalar field model in $3+1$ dimensions describing two $Q$-balls interacting
via a potential term. The Lagrangian reads:
\begin{equation}
\label{lag}
 {\cal L}=\frac{1}{2}\partial_{\mu} \Phi_1 \partial^{\mu} \Phi_1^*+
\frac{1}{2}\partial_{\mu} \Phi_2 \partial^{\mu} \Phi_2^* - V(\Phi_1,\Phi_2)
\end{equation}
where both $\Phi_1$ and $\Phi_2$ are complex scalar fields. The potential reads:
\begin{equation}
U(\Phi_1,\Phi_2)=\sum_{i=1}^2\left(
\alpha_i \vert\Phi_i\vert^6 - \beta_i \vert\Phi_i\vert^4 +
\gamma_i \vert\Phi_i\vert^2 \right)
+\lambda \vert\Phi_1\vert^2 \vert\Phi_2\vert^2
\end{equation} 
where $\alpha_i$, $\beta_i$, $\gamma_i$, $i=1,2$ are the standard potential parameters for each $Q$-ball, while $\lambda$ denotes the interaction parameter.

In \cite{vw} it was argued that a $\Phi^6$-potential is necessary in order to have classical
$Q$-ball solutions. This is still necessary for the model we have defined here,
since we want $\Phi_1=0$
and $\Phi_2=0$ to be a local minimum of the potential. A pure $\Phi^4$-potential which is bounded from below wouldn't fulfill these criteria. 
 
The Lagrangian (\ref{lag}) is invariant under the two global U(1) transformations
\begin{equation}
 \Phi_1 \rightarrow \Phi_1 e^{i\alpha_1} \ \ \ , \ \ \ 
\Phi_2 \rightarrow \Phi_2 e^{i\alpha_2} 
\end{equation}
which can be applied separately or together. As such the total conserved Noether
current $j^{\mu}_{(tot)}$, $\mu=0,1,2,3$, associated to these symmetries is just the sum of
the two individually conserved currents $j^{\mu}_{1}$ and $j^{\mu}_2$ with
\begin{equation}
j^{\mu}_{(tot)}= j^{\mu}_1 +j^{\mu}_2
 = -i \left(\Phi_1^* \partial^{\mu} \Phi_1 - \Phi_1 \partial^{\mu} \Phi_1^*\right)
-i  \left(\Phi_2^* \partial^{\mu} \Phi_2 - \Phi_2 \partial^{\mu} \Phi_2^*\right)\ \ .
\end{equation}
with $\partial_{\mu} j^{\mu}_{1}=0$, $\partial_{\mu} j^{\mu}_{2}=0$ and  $\partial_{\mu} j^{\mu}_{(tot)}=0$.

The total Noether charge $Q_{(tot)}$ of the system is then the sum of the two individual Noether charges $Q_1$ and $Q_2$:
\begin{equation}
 Q_{(tot)}=Q_1+Q_2= -\int j_1^0 d^3 x  -\int j_2^0 d^3 x
\end{equation}

Finally, the energy-momentum tensor reads:
\begin{equation}
T_{\mu\nu}=\sum_{i=1}^2 \left(\partial_{\mu} \Phi_i \partial_{\nu} \Phi_i^*
+\partial_{\nu} \Phi_i \partial_{\mu} \Phi_i^*\right) -g_{\mu\nu} {\cal L}
\end{equation}

\subsection{Ansatz}

We choose as Ansatz for the fields in spherical coordinates:
\begin{equation}
\label{ansatz1}
\Phi_i(t,r,\theta,\varphi)=e^{i\omega_i t+ik_i\varphi} \phi_i(r,\theta) \  \ ,
\ i=1,2
\end{equation}
where the $\omega_i$ and the $k_i$ are constants. Since we require $\Phi_i(\varphi)=\Phi_i(\varphi+2\pi)$, $i=1,2$, we have that $k_i\in \mathbb{Z}$.
It was moreover demonstrated in \cite{vw,kk} that $Q$-balls exist only in a specific
parameter range $\omega_{min} < \omega < \omega_{max}$ and that the charge $Q$
tends to infinity when either $\omega \rightarrow \omega_{min}$ or
$\omega \rightarrow \omega_{max}$. We discuss the limits in the 2 $Q$-ball system
in the following section.

The Noether charges of the solution then read:

\begin{equation}
Q_i = 2\omega_i \int \vert \Phi_i \vert^2 \ d^3 x = 4\pi \omega_i \int_{0}^{\pi} \int_0^{\infty}
r^2 \sin\theta \ dr \ d\theta \ \phi_i^2  \ \ \ , \ \ i=1,2
\end{equation}
while the energy is given by the volume integral of the $tt$-component of the energy-momentum
tensor:
\begin{eqnarray}
E=\int T_{00} \  d^3 x &=&2\pi \int\limits_0^{\pi}\int\limits_0^{\infty}\left[ r^2 
\ \sin\theta \ dr \ d\theta \sum_{i=1}^2\left(\omega_i \phi_i + (\phi_i')^2 
+ \frac{(\dot{\phi}_i)^2 }{r^2} \right.\right. \nonumber \\    
&+& \left. \left. \frac{k_i^2\phi_i^2}{r^2\sin^2\theta} +
\alpha_i \vert\phi_i\vert^6 -\beta_i \vert\phi_i\vert^4 +
\gamma_i \vert\phi_i\vert^2 \right)
+ \lambda \vert\phi_1\vert^2 \vert\phi_2\vert^2 \right]
\end{eqnarray}
where the prime and dot denote the derivative with respect to $r$ and $\theta$, 
respectively.

For $k_i\neq 0$, the solutions have non-vanishing angular momentum that is quantised.
The total angular momentum $J$ is the sum of the angular momenta
of the two individual $Q$-balls:
\begin{equation}
J=\int T_{0\varphi} d^3 x =J_1+J_2= k_1 Q_1 + k_2 Q_2
\end{equation}
We will thus in the following refer to solutions with $k_i=0$ as non-spinning
and to solutions with $k_i\neq 0$ as spinning.

The Euler-Lagrange equations read:

\begin{equation}
\label{eom}
\phi_i''+ \frac{2}{r} \phi_i' + \frac{1}{r^2} \ddot{\phi}_i
+ \frac{1}{r^2} \cot\theta \dot{\phi}_i - \frac{k_i^2}{r^2\sin^2\theta}\phi_i  
+ \omega_i^2 \phi_i = 3\alpha_i \phi_i^5 - 2\beta_i \phi_i^3 + \gamma_i \phi_i
+ \lambda \phi_i \phi_k^2 
\end{equation}
with $i=1,2$ and $k\neq i$.

The boundary conditions, which result from requirements
of regularity, finiteness of the energy and the symmetry of the
solution, are:
\begin{equation}
\label{bc1}
\partial_r\phi_i(r=0,\theta)= 0 \ , \  
\partial_{\theta}\phi_i(r=\infty,\theta)=0 \ , \   
\partial_{\theta}\phi_i(r,\theta=0,\pi)=0  \ , \ i=1,2  \ .
\end{equation}
for non-spinning solutions with $k_i=0$ and 
\begin{equation}
\label{bc2}
\phi_i(r=0,\theta)= 0 \ , \  
\phi_i(r=\infty,\theta)=0 \ , \   \phi_i(r,\theta=0,\pi)=0  \ , \ i=1,2  \ .
\end{equation}
for spinning solutions $k_i\neq 0$.

\subsection{Bounds on $\omega_1$ and $\omega_2$ in the 2-$Q$-ball system}
In \cite{vw,kk} the bounds on the frequency $\omega$ have been discussed in the case of
one $Q$-ball. Here, we note that these bounds have to be modified if one considers two interacting $Q$-balls.
The set of equations (\ref{eom}) can be interpreted as the mechanical equations describing
the frictional motion of a particle in two dimensions. The effective potential in this case reads:
\begin{equation}
V(\phi_1,\phi_2)= \frac{1}{2} (\omega_1^2\phi_1^2 + \omega_2^2\phi_2^2) - \frac{1}{2} U(\phi_1,\phi_2)
\end{equation}
$Q$-ball solutions exist provided the configuration $(\phi_1=0$,$\phi_2=0)$ corresponds to
a local maximum of the effective potential and provided the effective potential
has positive values in any radial direction from the origin in the $\phi_1-\phi_2-$plane.
This leads to non-trivial bounds for the parameters $\omega_1$ and $\omega_2$.

The former condition leads to the requirement that
\begin{equation}
\omega_1^2 < \omega_{1,max}^2 = \gamma_1 \ \ , \ \  
 \omega_2^2 < \omega_{2,max}^2 = \gamma_2 \ \ . 
 \end{equation}
 
 The latter condition leads to a more complicated domain of existence in the 
 $\omega_1$-$\omega_2$-plane. To describe this condition, we introduce the polar 
 decomposition of $\phi_1$ and $\phi_2$ as follows:
 \begin{equation}
 \phi_1=\rho\cos\chi \ \ , \ \ \phi_2=\rho\sin\chi \ .
 \end{equation}
 where $0 \le \chi < 2\pi$ and $0 \le \rho < \infty$.

 The condition on the frequencies $\omega_1$ and $\omega_2$ then  read:
 \begin{equation}
 \omega_1^2 \cos^2\chi + \omega_2^2 \sin^2\chi > 
 (\omega_1^2 \cos^2\chi + \omega_2^2 \sin^2\chi)_{min} = {\rm min}_{\rho}[U(\rho,\chi)/\rho^2]   \ \ , \ \ \forall \ \chi
 \end{equation}
 
 In the particular case that we have studied throughout this paper, namely $\alpha_1=\alpha_2=1$,
 $\beta_1=\beta_2=2$ and $\gamma_1=\gamma_2=1.1$
  this inequality takes the form:
 \begin{eqnarray}
  \omega_1^2 \cos^2\chi + \omega_2^2 \sin^2\chi & > &  \left[-5\lambda^2 \cos^4\chi\sin^4\chi+
  20\lambda\cos^2\chi\sin^2\chi (\cos^4\chi + \sin^4\chi) \nonumber \right. \\
  &+& \left.
  2(\cos^8\chi + \sin^8\chi + 11 \cos^6\chi\sin^2\chi + 11 \sin^6\chi\cos^2\chi \nonumber \right. \\
  &-& \left.
  20 \cos^4\chi\sin^4\chi)\right]/(\cos^4\chi +\sin^4\chi -\cos^2\chi\sin^2\chi)
 \end{eqnarray}
 
 For $\chi= n \pi/2$, $n=0,1,2,...$, we recover the results of the one $Q$-ball system
 discussed in \cite{vw,kk}. For all other values of $\chi$, the limiting values for
 $\omega_1$ and $\omega_2$ will depend on the value of the interaction coupling $\lambda$.
 E.g. for $\phi_1=\phi_2$, i.e. $\chi=\pi/4$, we find :
 
 \begin{equation}
 \omega_1^2 + \omega_2^2  > 1/5 + \lambda - 1/8 \lambda^2
 \end{equation}
 Thus, for small $\lambda$, the lower bound on the value of 
 $\omega_1^2 + \omega_2^2$ will be larger than in the non-interacting limit.

\section{New non-spinning $Q$-ball solutions for $\alpha_2=\beta_2=\gamma_2=\lambda=0$}
In order to be able to understand the structure of a system of two $Q$-balls, 
we have reconsidered
the one $Q$-ball system. We set all quantities with index ``2'' to zero in the following
and omit the index ``1'' for the remaining quantities. 

In this section, we would like to point 
out that more than the previously in the literature
discussed solutions exist.

For this, we first consider the equation for one $Q$-ball 
with vanishing potential. This reads:
\begin{equation}
\phi''+ \frac{2}{r}\phi' + \frac{1}{r^2} \ddot{\phi}
+ \frac{1}{r^2} \cot\theta \dot{\phi} - \frac{k^2}{r^2\sin^2\theta}\phi   
+ \omega^2 \phi=0
\end{equation}
Although the solutions of the above equation are 
well known, it will be useful for the following to recall their properties.
Using the standard separation of variables, the solutions read:
\begin{equation}
\phi(r,\theta,\varphi)\propto \frac{J_{L+1/2}(\omega r)}{\sqrt{r}} Y_L^k(\theta,\varphi)
\end{equation}

where $J$ denotes the Bessel function, while $Y_L^k$ 
are the standard spherical harmonics with $-L \le k \le L$.

One may hope that solutions of the full equations with the discrete symmetries corresponding to the ones
of the spherical harmonics will exist. Of course, the non-linear potential interaction will
deform the radial part of the solutions of the linear equation in a highly non-trivial manner.

The solutions of the full equation constructed so far for $k=0$ have been spherically symmetric.
With the above arguments, axially symmetric solutions should equally exist with an angular dependence
of the form $Y_L^0$, e.g. for $L=1$, the angular dependence should be 
of the form $\cos\theta$. In the following, we will denote the solutions of the full
non-linear equations with angular symmetries corresponding to the symmetries of the spherical harmonic $Y_L^k$
by $\phi_L^k$.

\subsection{Numerical results}

\begin{figure}
\includegraphics[width=8cm]{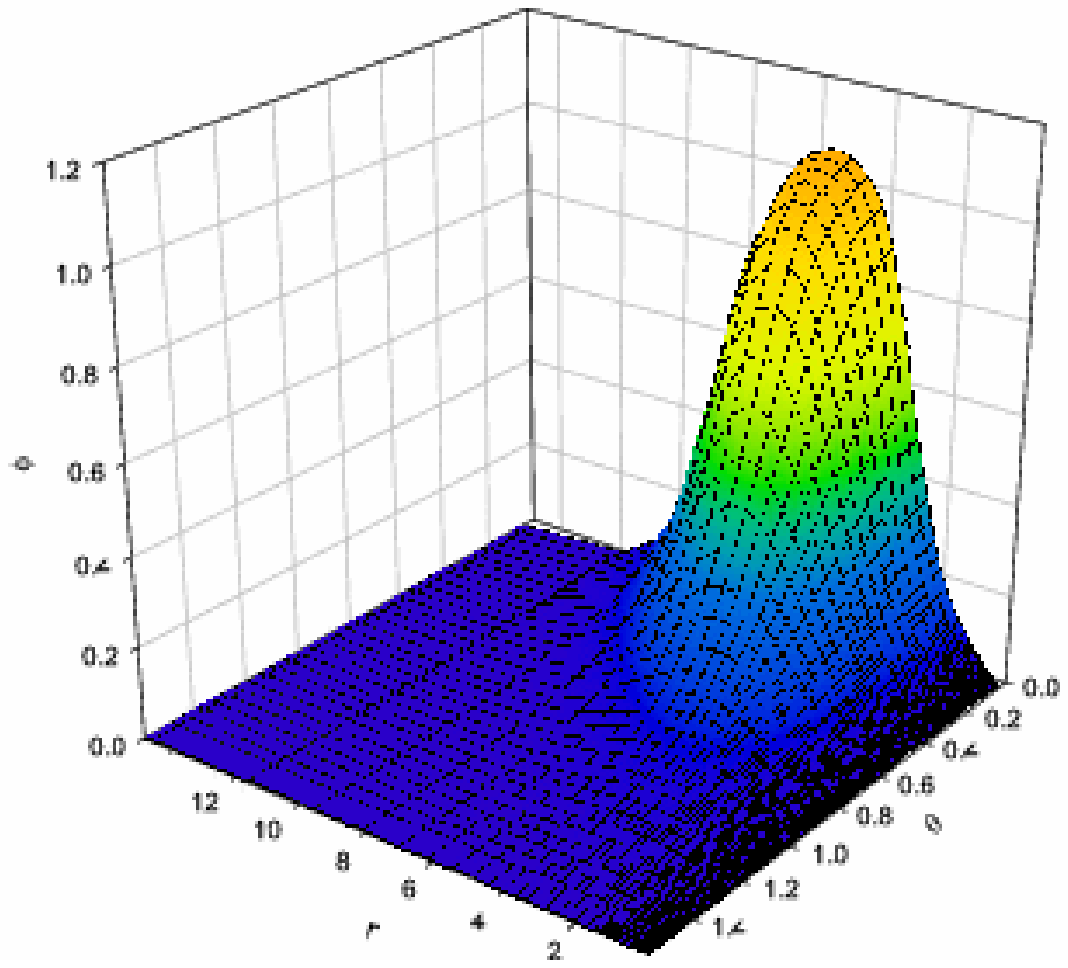} 
\includegraphics[width=8cm]{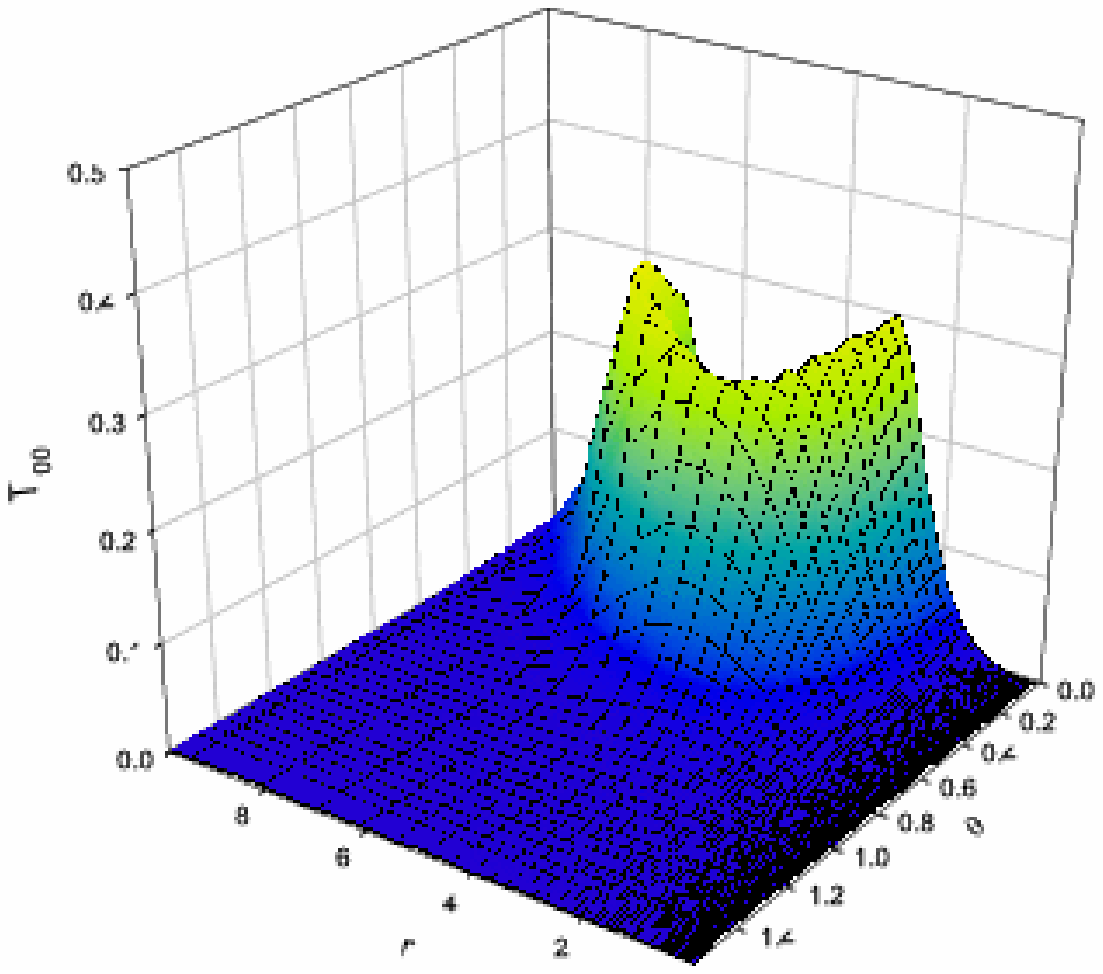}
\caption{\label{fig1a} The profile of the function $\phi_1^0$ is shown for
$\omega = 0.8$, $\alpha=1$, $\beta=2$, $\gamma=1.1$ (left).
The corresponding energy density $T_{00}$ is also given (right).}
\end{figure}

The partial differential equation has been solved numerically
subject to the boundary conditions (\ref{bc1}) or (\ref{bc2}) using the finite difference solver FIDISOL
\cite{fidi}.
We have mapped the infinite interval of the $r$ coordinate $[0:\infty]$ to the
finite
compact interval $[0:1]$ using the new coordinate $z:=r/(r+1)$.  
We have typically used grid sizes of $150$ points in $r$-direction and $50$
points in $\theta$ direction. The solutions have relative errors of $10^{-3}$
or smaller.
Throughout this section, we choose $\alpha_1\equiv\alpha=1$,
$\beta_1\equiv\beta=2$, $\gamma_1\equiv\gamma=1.1$.

In Fig.\ref{fig1a} (left), we show the profile of a new solution that we obtained for $k=0$ and $\omega=0.8$. This solution
looks like a deformation of the spherical harmonic $Y_1^0$ with the appropriate symmetry with respect to $\theta=\pi/2$ and is clearly axially symmetric. In particular it fulfills $\phi_1^0(r,\pi/2)=0$.



The field $\phi_1^0(r,\theta)$ is maximal at a finite distance from the origin
on the positive $z$-axis. Moreover, the configuration is anti-symmetric under
reflexion through the $x-y-$plane, i.e. under $\theta \rightarrow \pi-\theta$. Thus the solution
is parity-odd: $P=-1$.
Note that we have only plotted the function for $\theta\in [0:\pi/2]$, but that
we have verified the symmetry of the solution.

\begin{figure}[!htb]
\centering
\leavevmode\epsfxsize=11.0cm
\epsfbox{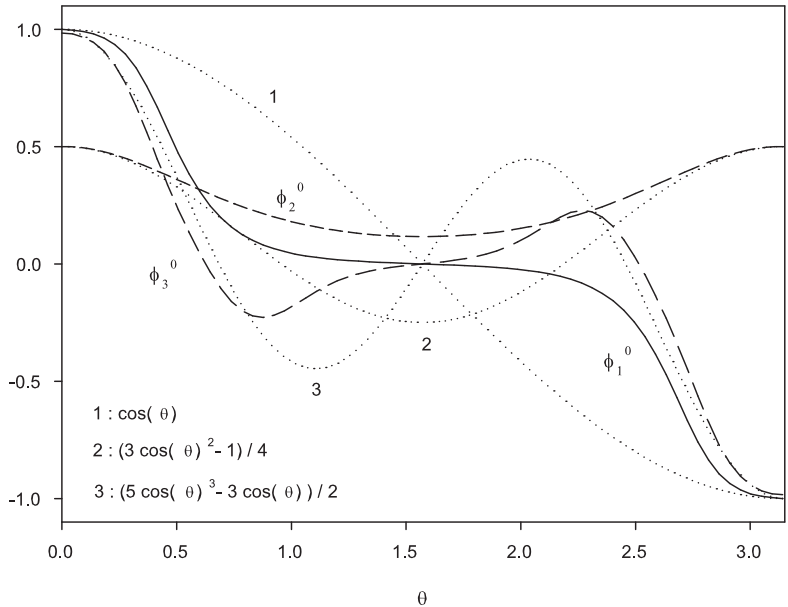}\\
\caption{\label{fignew}  $\phi_1^0$ (solid), $\phi_2^0$ (short dashed) and $\phi_3^0$ (long dashed) are shown as functions
of $\theta$ for fixed value of $r$. We have chosen $r\sim 5 $ for $\phi_1^0$, $r\sim 2$ 
for $\phi_2^0$ and $r\sim 6 $ for $\phi_3^0$. The
corresponding spherical harmonics $Y_1^0\propto \cos(\theta)$, $Y_2^0\propto 3\cos(\theta)^2-1$ 
and $Y_3^0\propto 5\cos(\theta)^3-3\cos\theta$ 
(with an appropriate normalisation) are also shown.  }
\end{figure}

We also present the corresponding energy density $T_{00}$ in Fig.\ref{fig1a} (right). It shows that the density of the solution
 is mainly  concentrated within two small ``balls'' situated around  the positive
$z$-axis (at $z\approx 2.4$ and $z\approx 7.6$) and separated by a minimum (at $z\approx 5$). 
The position of this minimum coincides with the maximum of the scalar
field $(\phi_1^0)_{max} \approx \phi_1^0(5,0) \approx 1.2$. It can be checked that this value corresponds roughly to a local minimum
of the potential while the partial derivatives are evidently small in this region,
explaining the occurrence of a minimal value of the energy density at $(x,y,z) \approx (0,0,5)$. 
Of course, due to the anti-symmetry of the solution this pattern is equally given on the negative
$z$-axis. 

\begin{figure}[!htb]
\centering
\leavevmode\epsfxsize=10.0cm
\epsfbox{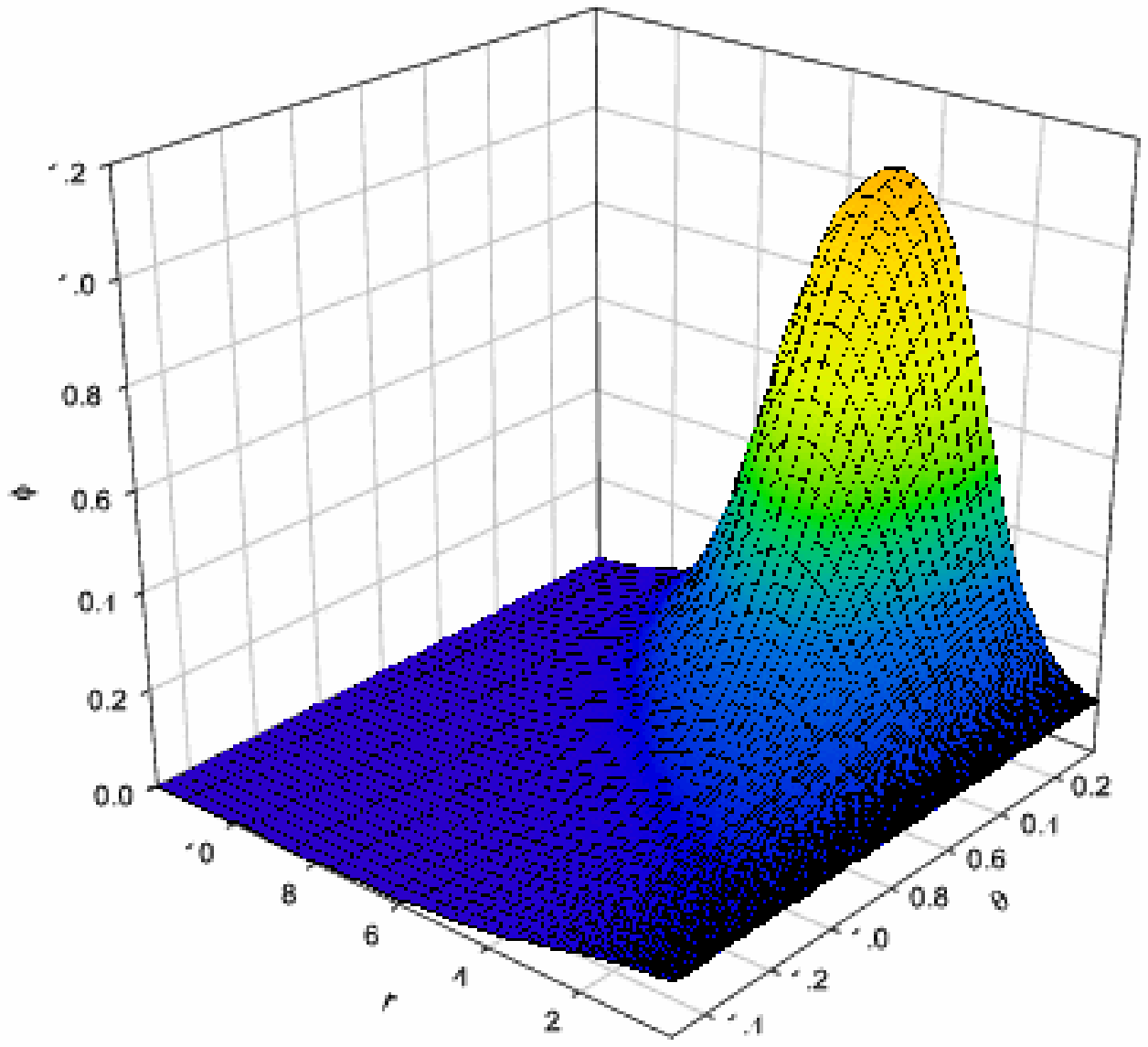}\\
\caption{\label{fig2} The profile of the function $\phi_2^0$  for
$\omega = 0.8 $, $\alpha=1$, $\beta=2$, $\gamma=1.1$ }
\end{figure}

The classical energy and charge of this new solution is higher
than that of the spherically symmetric $k=0$ solution (see Table 1 and Table 2 below), however lower than that
of the $k=1$ spinning $Q$-ball. 

In order to investigate further our idea of constructing  new solutions as deformations
of the spherical harmonics, we have also investigated solutions with higher value of $L$
and we  managed to  construct
solutions $\phi_2^0$ and $\phi_3^0$ corresponding in their angular symmetries to those of the spherical harmonics $Y_2^0 \propto 3\cos^2\theta -1$ and $Y_3^0 \propto 5\cos^3\theta -3\cos\theta$, respectively.

In Fig.\ref{fignew}, we plot $\phi_1^0$, $\phi_2^0$ and $\phi_3^0$
as functions of $\theta$ for a fixed value of $r$ together with the corresponding spherical harmonics
$Y_1^0$, $Y_2^0$ and $Y_3^0$. Here, we have chosen $r\sim 5$ for $\phi_1^0$, $r\sim 2$ for $\phi_2^0$
and  $r\sim 6$ for $\phi_3^0$. The first thing to notice is that
the symmetries of the solutions  $\phi_1^0$, $\phi_2^0$ and $\phi_3^0$ with respect to reflection
at $\theta=\pi/2$ are exactly equal to those of the corresponding spherical harmonics. The actual solutions are, of course, deformed with respect
to the spherical harmonics, but the correspondence is apparent. 
E.g. the solution $\phi_2^0$ has $\partial_\theta \phi_2^0(r, \pi/2)=0$
(in contrast to the solution $\phi_1^0$ which has $\phi_1^0(r, \pi/2)=0$). We don't show the energy
density of $\phi_2^0$ and $\phi_3^0$ here, since it resembles that shown in Fig. 1. 

We believe that
the correspondence also holds for higher spherical harmonics.

\begin{table}
\begin{center}
\begin{tabular}{|c||c|c|c|c|}
\hline\hline
 & $\phi_0^0$ & $\phi_1^0$ & $\phi_2^0$ & $\phi_1^1$\\
\hline
$E$ & $73.6$ & $141.6$  & $170.9$ & $223.5$\\
$Q/\omega$ & $75.2$ & $146.9$ & $176.1$ & $220.4$ \\
$P$ & $+1$  & $-1$ & $+1$  & $+1$ \\
symmetry & spherical & axial & axial  & axial \\
\hline
\end{tabular}
\caption{The energy $E$, the charge per frequency $Q/\omega$, the parity $P$
and the symmetry of the first few $Q$-ball solutions is given for
$\omega=0.8$. }
\end{center}
\end{table}

\begin{table}
\begin{center}
\begin{tabular}{|c||c|c|c|c|}
\hline\hline
 & $\phi_0^0$ & $\phi_1^0$ & $\phi_2^0$ & $\phi_1^1$\\
\hline
$E$ & $61.2$ & $115.8$  & $179$ & $195.6$\\
$Q/\omega$ & $60.0$ & $114.7$ & $192.1$ & $186.3$ \\
$P$ & $+1$  & $-1$ & $+1$  & $+1$ \\
symmetry & spherical & axial & axial  & axial \\
\hline
\end{tabular}
\caption{The energy $E$, the charge per frequency $Q/\omega$, the parity $P$
and the symmetry of the first few $Q$-ball solutions is given for
$\omega=0.84$. }
\end{center}
\end{table}

Since we have presented strong numerical evidence that the correspondence with
the spherical harmonics holds, it is justified to label 
the different solutions of the field equation
by means of the quantum numbers of the corresponding spherical harmonic, i.e. by $L$ and $k$ 
referring to
$Y_L^k$, with $L,k$ integers and $-L \le k \le L $. Needless to say that the numerical construction
becomes more involved when the difference $L-|k|$ increases.
Adopting these notations and fixing the potential according to $\alpha=1$, $\beta=2$, $\gamma=1.1$,
 we find for the solutions corresponding to $\omega = 0.8$ and $\omega=0.84$
the values for the energy $E$ and  charge per frequency $\frac{Q}{\omega}$ given in Table 1 and Table 2, respectively.


The first three solutions $\phi_L^0$, $L=0,1,2$ in this list are static (i.e. non spinning) while the last, $\phi_1^1$, is stationary (i.e. spinning). 
For all the solutions we constructed,  the energy of the non-spinning solutions is lower that the energy of the spinning ones.

\begin{figure}[!htb]
\centering
\leavevmode\epsfxsize=12.0cm
\epsfbox{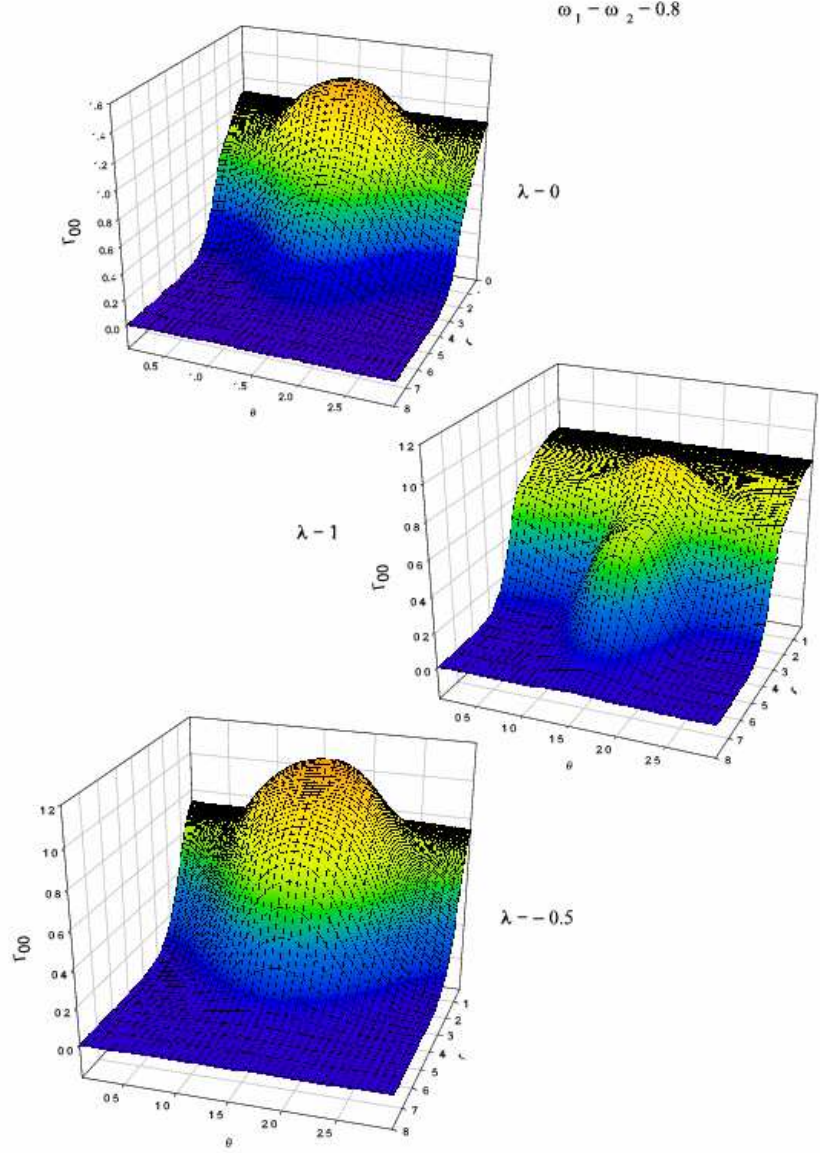}\\
\caption{\label{t00} The energy density $T_{00}$ of the 2-$Q$-ball solution
consisting of a spherically symmetric, non-spinning $Q$-ball ($k_1=0$)
and a spinning $Q$-ball ($k_2=1$) is shown for
$\omega_1 = \omega_2 = 0.8$, $\alpha_1=\alpha_2=1$,
$\beta_1=\beta_2=2$, $\gamma_1=\gamma_2=1.1$ and
for three different values of $\lambda = 0$, $1$, $-0.5$.   }
\end{figure}

\section{Interacting $Q$-balls}
Since in supersymmetric extensions of the Standard Model, $Q$-balls exist
that result from the interaction of several scalar fields, we here
investigate the interaction of two classical $Q$-balls as toy-model for these
systems. 

For two spherically symmetric $Q$-balls ($k_1=k_2=0$) in interaction, 
the 2-$Q$-ball solution is still spherically symmetric and the domain
of existence in the $\omega_1$-$\omega_2$-plane can be  determined
by using the reasoning given in Section 2.2. Here, we put the emphasis
on solutions where the two $Q$-balls have different symmetries and study the effect
of the direct interaction parameterized by the coupling constant $\lambda$.

We believe that a particularly interesting case is the interaction between
a spherically symmetric, non-spinning $Q$-ball ($k_1=0$) and a spinning $Q$-ball ($k_2=0$).
We have thus restricted our analysis to this case and set $k_1=0$ and $k_2=1$ in the following.

Note that we will index all quantities related to the spherical $Q$-ball in the following
with ``1'', while all quantities related to the axially symmetric $Q$-ball will be indexed
with ``2''.

For later use, we define the ``binding energy'' of the solution
 according to 
 \be
 \Delta E = E- E_{k_1=0}- E_{k_2=1} \ . 
 \ee
 It represents 
 the difference between the energy $E$ of the
2-$Q$-ball configuration and the sum of the energies
of the two single (i.e. non-interacting) $Q$-balls $E_{k_1=0}$, $E_{k_2=1}$ with the same frequency. We expect that those solutions which have $\Delta E <0$ to be stable,
while those with $\Delta E > 0$ would be unstable. 
\subsection{Numerical results}

We have solved the two coupled partial differential equations
using the solver FIDISOL \cite{fidi} for several values of $\omega_1$, $\omega_2$  and $\lambda$
and fixing $\alpha_1=\alpha_2=1$, $\beta_1=\beta_2=2$ and $\gamma_1=\gamma_2=1.1$.
As starting profiles, we have used the corresponding non-interacting $Q$-ball solutions.
For $\lambda=0$, these solve the two decoupled partial differential equations. We have
then slowly increases the parameter $\lambda$ to obtain the interacting solutions.
 
\subsubsection{$\omega_1=\omega_2$}
In order to understand the influence of the interaction parameter $\lambda$, we show the
energy density $T_{00}$ for $\omega_1=\omega_2=0.8$ and three different values 
of $\lambda$ in Fig.\ref{t00}.
For $\lambda=0$, the two $Q$-balls are non-interacting and the energy density is just a simple
superposition of the energy densities of the two individual $Q$-balls.
For $\lambda \neq 0$ the $Q$-balls interact.
For $\lambda > 0$, it is energetically favourable to have the two $Q$-balls' cores
in different regions of space. As seen in Fig.\ref{t00} for $\lambda=1$, the  
spinning $Q$-ball
seems to be ``pushed away'' from the non-spinning, spherically symmetric
one. For $\lambda < 0$, it is energetically favourable to have two $Q$-balls sitting
``on top of each other''. This is shown in Fig.\ref{t00} for $\lambda=-0.5$, where the two $Q$-balls
seem to be localised at the same place.

\begin{figure}[!htb]
\centering
\leavevmode\epsfxsize=10.0cm
\epsfbox{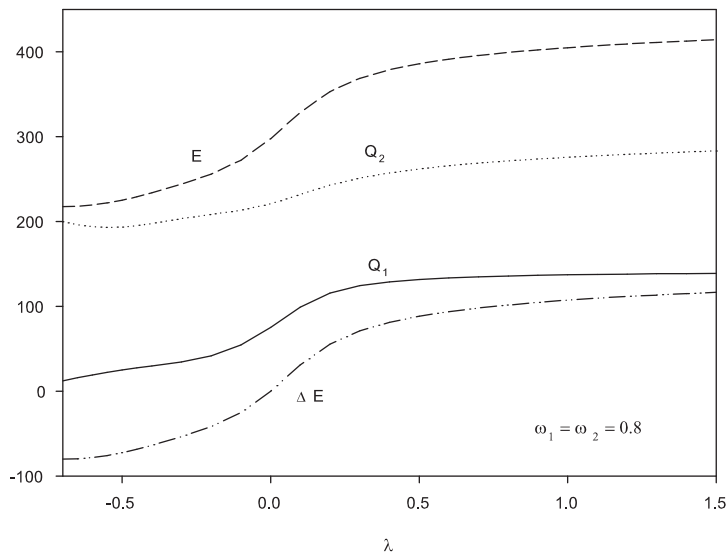}\\
\caption{\label{lambdavary} The quantities $E$, $Q_1$, $Q_2$ and $\Delta E$ are shown
as functions of the interaction parameter $\lambda$ 
with $\omega_1= \omega_2= 0.8$, $\alpha_1=\alpha_2=1$,
$\beta_1=\beta_2=2$, $\gamma_1=\gamma_2=1.1$    }
\end{figure}

We have also studied the dependence of the energy $E$, the binding energy $\Delta E$ and the two charges
$Q_1$ and $Q_2$ on the interaction parameter $\lambda$. The results are shown in Fig.\ref{lambdavary}
for $\omega_1=\omega_2=0.8$. All quantities increase with the increase of $\lambda$, specifically
it is evident that
the 2-$Q$-ball configuration is energetically more stable for $\lambda <0$ than for $\lambda > 0$.  Specifically, we would thus expect the solution to be stable for
 $\lambda <0$ and unstable for $\lambda > 0$.

\begin{figure}
\includegraphics[width=8cm]{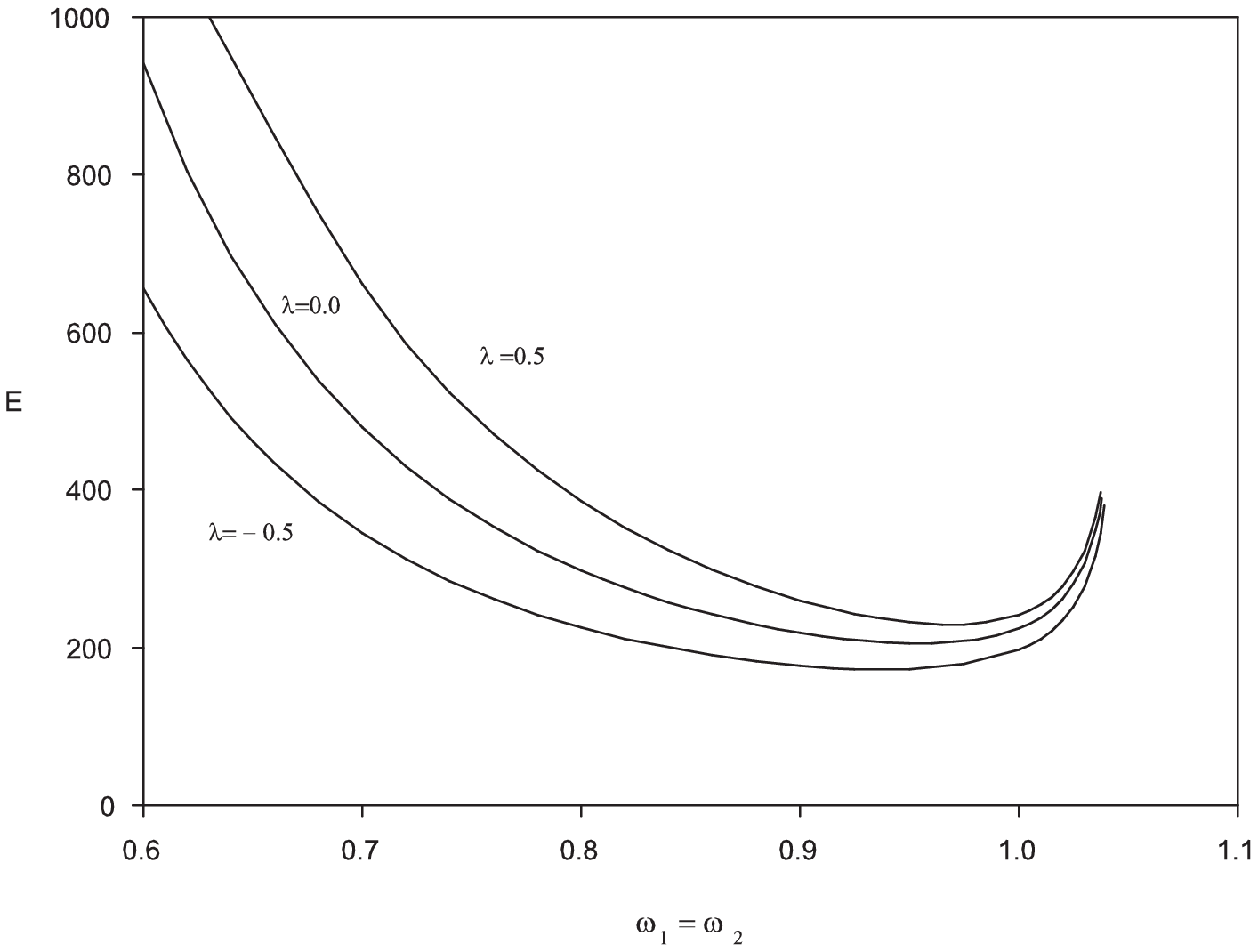}
\includegraphics[width=8cm]{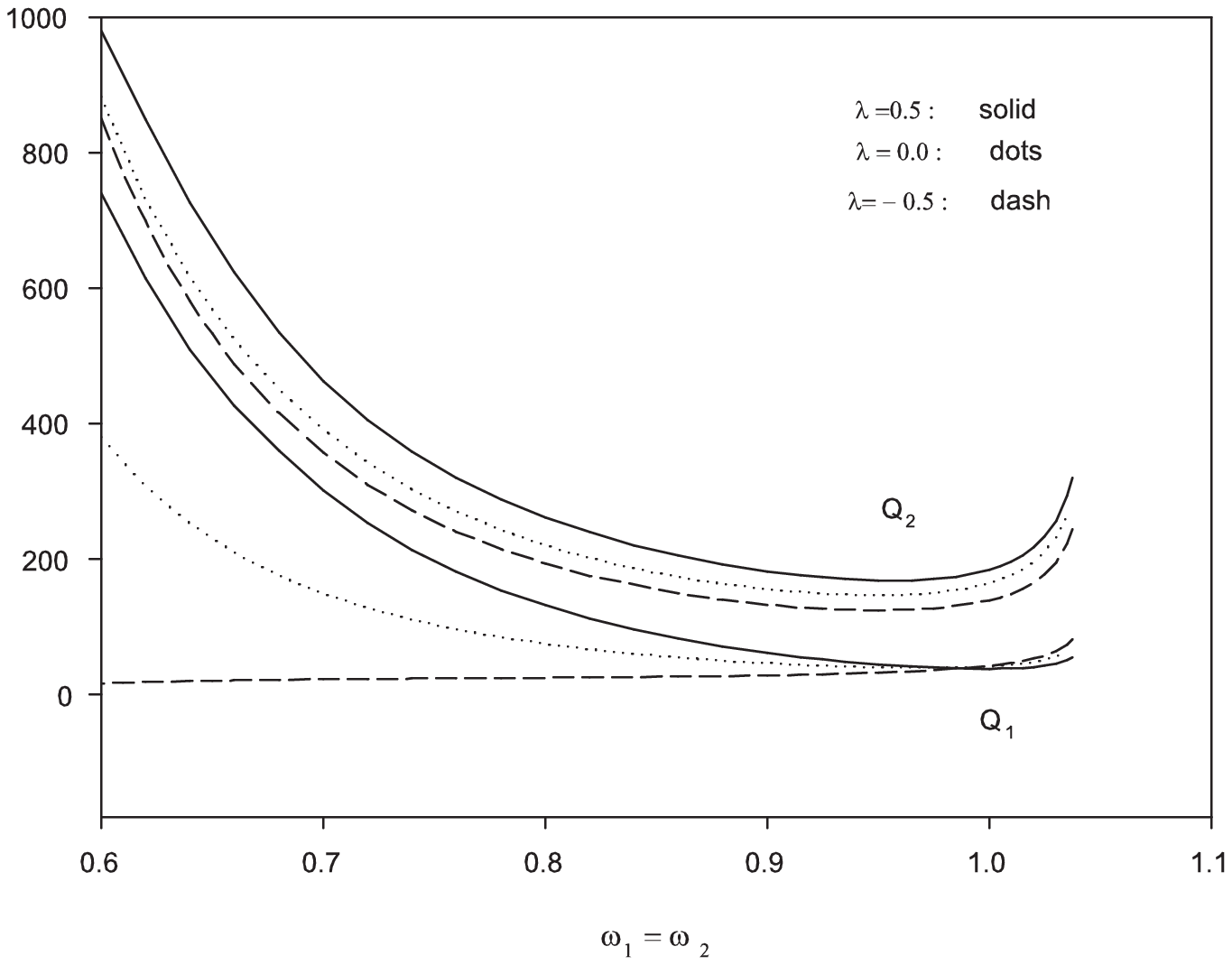}
\caption{\label{fig_qq2} The energy $E$ (left) and the charges $Q_1$ and $Q_2$ (right)
are shown as functions of the frequency $\omega_1=\omega_2$ 
for $\lambda= -0.5,0,0.5$.}
\end{figure}

Following our discussion in Section 2.2, we have also studied the dependence of the energy $E$ and
of the charges $Q_1$ and $Q_2$ on the frequencies $\omega_1$ and $\omega_2$. Our results for
$\omega_1=\omega_2$ are shown in Fig.\ref{fig_qq2} for $\lambda=-0.5$, $0$ and
$0.5$, respectively.

As expected, the energy $E$ for a given frequency $\omega_1=\omega_2$ is higher (resp. lower)
than in the non-interacting case for positive (resp. negative) values of $\lambda$.

As before, we find that the solutions exist in a given interval of the frequency:
$\omega_{1,min}(\lambda) \leq \omega \leq \omega_{1,max}(\lambda)$ (and equally for $\omega_2$ since $\omega_1=\omega_2$). 
We have determined the bounds on $\omega_1$ and $\omega_2$ in Section 2.2 for two spherically
symmetric $Q$-balls. Here, we would expect that these values change slightly since we
have a system of one spherically symmetric and one axially symmetric $Q$-ball. However,
we see that the qualitative results are similar here.  We observe that for $\lambda \ge 0$,
the values of the energy $E$ and of the charges $Q_1$, $Q_2$ diverge
at $\omega_1=\omega_{1,min}$ and 
$\omega_1=\omega_{1,max}$.
Following the discussion of Section 2.2 we find that the maximal
value of $\omega_1$ is independent of $\lambda$. This can be clearly seen in 
Fig. \ref{fig_qq2}
where the energy $E$ and the charges $Q_1$ and $Q_2$ diverge at $\omega_1=\omega_{1,max}\approx
1.035$ for  all three values of $\lambda$. Note that this maximal
value is only slightly lower than the bound given in Section 2.2: $\omega^2_{1,max}=1.1$. The reason why the bound is not equal is that here we are dealing with
an axially symmetric solution interacting with a spherically symmetric one.
Analytic arguments of the type done in Section 2.2 are, however, only possible
if the Euler-Lagrange equations are ordinary differential equations, i.e.
only in the case where the solutions are spherically symmetric. So, it is not
surprising that the analytic values differ from the numerical ones. 

On the other hand, the minimal value of $\omega_1$ is $\lambda$-dependent. This can be seen
in Fig. \ref{fig_qq2}. We have given our results only for $\omega \ge 0.6$ in this figure
since the construction of solutions becomes increasingly difficult for $\omega < 0.6$. However, it can be
clearly seen that the energy $E$ and $Q_2$ diverge at different values of $\omega_1=\omega_{1,min}$.
In agreement with Section 2.2., we find that $\omega_{1,min}$ is increasing for increasing (and
small) $\lambda$.

\begin{figure}[!htb]
\centering
\leavevmode\epsfxsize=10.0cm
\epsfbox{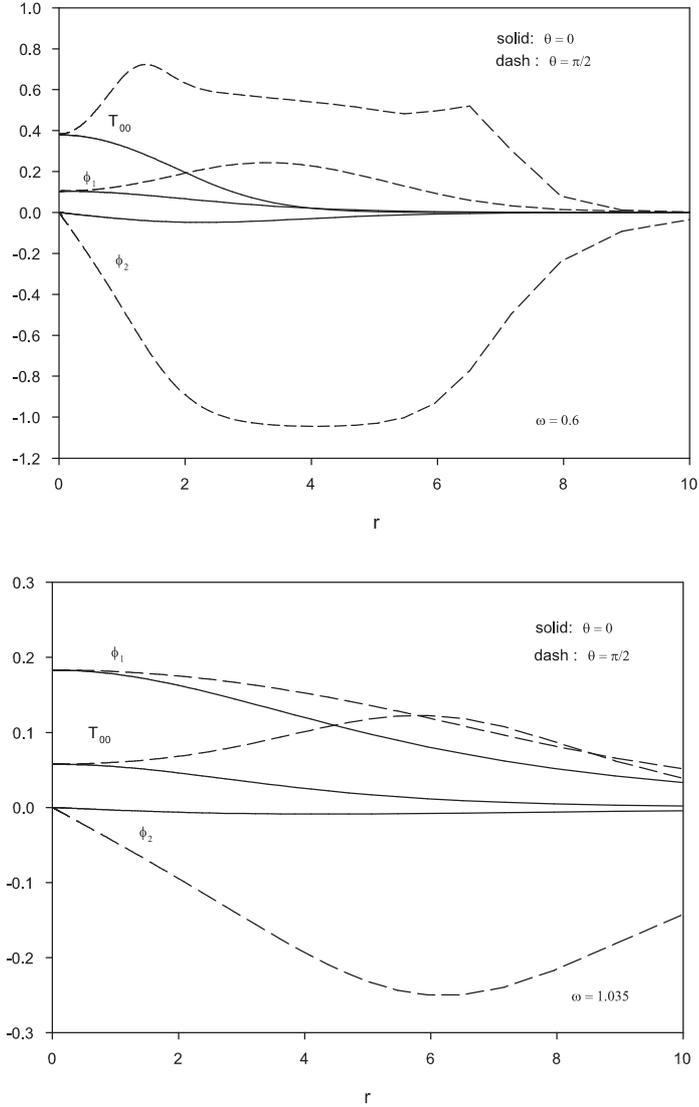}\\
\caption{\label{coupe} The profiles of $\phi_1$, $\phi_2$ and the energy density $T_{00}$
are shown for $\theta=0$ and $\theta=\pi/2$, respectively. Here $\lambda= -0.5$. The upper
figure is for $\omega=0.6$, the lower for $\omega=\omega_{1,max}\approx 1.035$.}
\end{figure}

For $\lambda < 0$ the behaviour at the lower bound of $\omega_1$ changes. We observe that $Q_1$ corresponding to the spherically symmetric field $\phi_1$ decreases when $\omega_1$ decreases.
The analysis of the profile of the solution reveals that the field $\phi_1$ deviates only slightly from
the spherically symmetric configuration for frequencies close to $\omega_{1,max}$. However, it 
gets more and more deformed in the
equatorial plane when $\omega_1$ decreases. At the same time, 
the field $\phi_2$ increases in the equatorial plane.
This phenomenon is illustrated in Fig.\ref{coupe} for $\lambda = - 0.5$, $\omega = 0.6$ and $\omega= \omega_{1,max}$, respectively.
In this figure, the fields $\phi_1$ and $\phi_2$ as well as the energy density $T_{00}$ are shown as function of $r$ for two angles
$\theta= 0$ and $\theta = \pi/2$, respectively. 



\subsubsection{Solutions with $\omega_1 \neq \omega_2$}
\begin{figure}[!htb]
\centering
\leavevmode\epsfxsize=10.0cm
\epsfbox{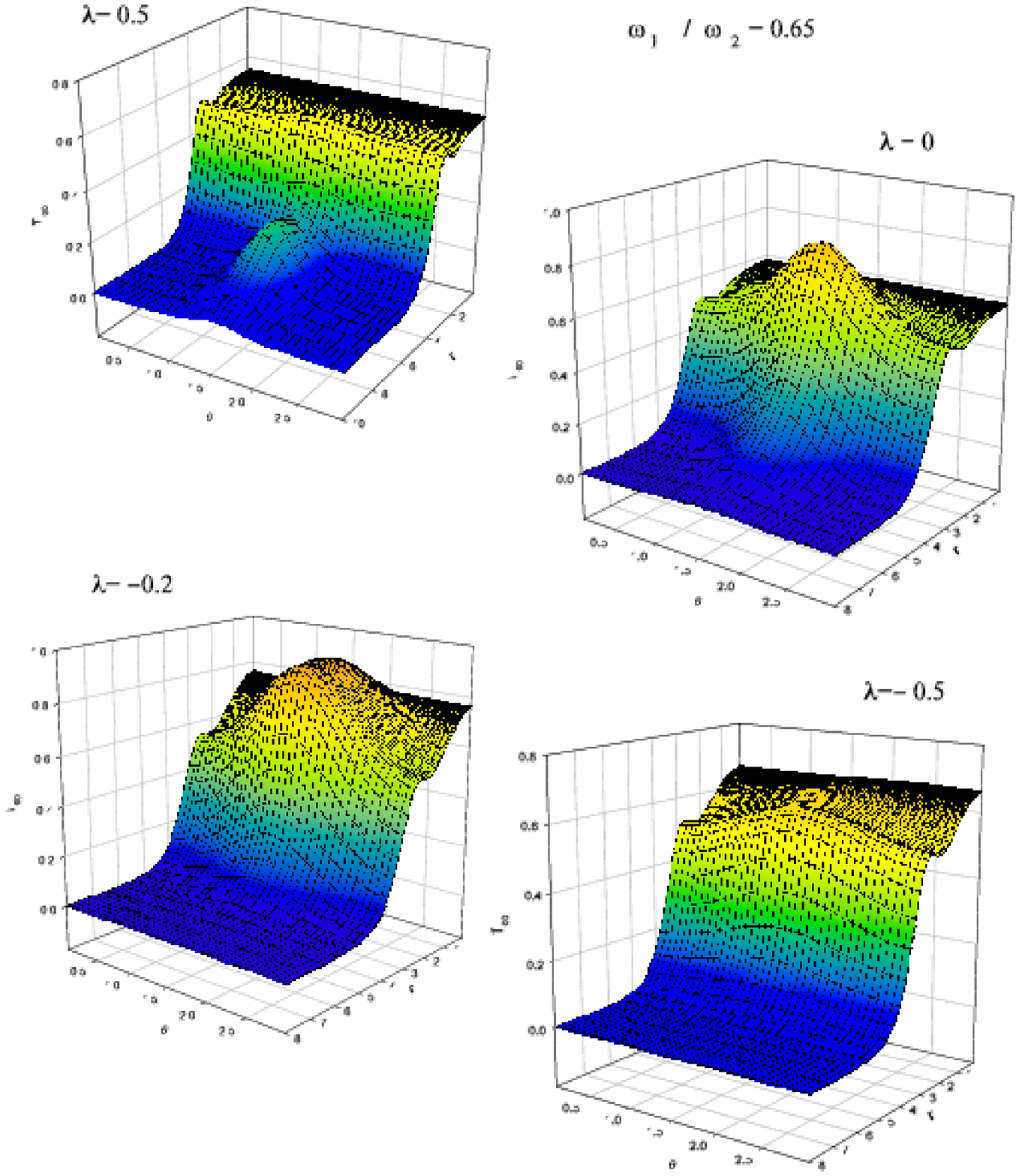}\\
\caption{\label{t00bis} The energy density $T_{00}$ of the 2-$Q$-ball solution
consisting of a spherically symmetric, non-spinning $Q$-ball ($k_1=0$)
and a spinning $Q$-ball ($k_2=1$) is shown for
$\omega_1 = 0.65$, $\omega_2 = 1$, $\alpha_1=\alpha_2=1$,
$\beta_1=\beta_2=2$, $\gamma_1=\gamma_2=1.1$ and
for four different values of $\lambda = 0.5$, $0$, $-0.2$ and $-0.5$.}
\end{figure}

We have also constructed 2-$Q$-ball solutions for
 $\omega_1 \neq \omega_2$. 
The energy density $T_{00}$ of a 2-$Q$-ball solution corresponding 
to  $\omega_1=0.65$ and $\omega_2=1$
is shown in Fig.\ref{t00bis} for four different values of $\lambda$.
The result is qualitatively similar to the case $\omega_1=\omega_2$.
This figure however suggests very clearly that for $\lambda < 0$ 
the $k_2=1$ $Q$-ball has tendency to disappear from the 2-$Q$-ball system.
For instance the maximal value of the $\phi_2$ field, $\vert \phi_{2,max}\vert$, 
decreases for decreasing $\lambda$.

\begin{figure}[!htb]
\centering
\leavevmode\epsfxsize=12.0cm
\epsfbox{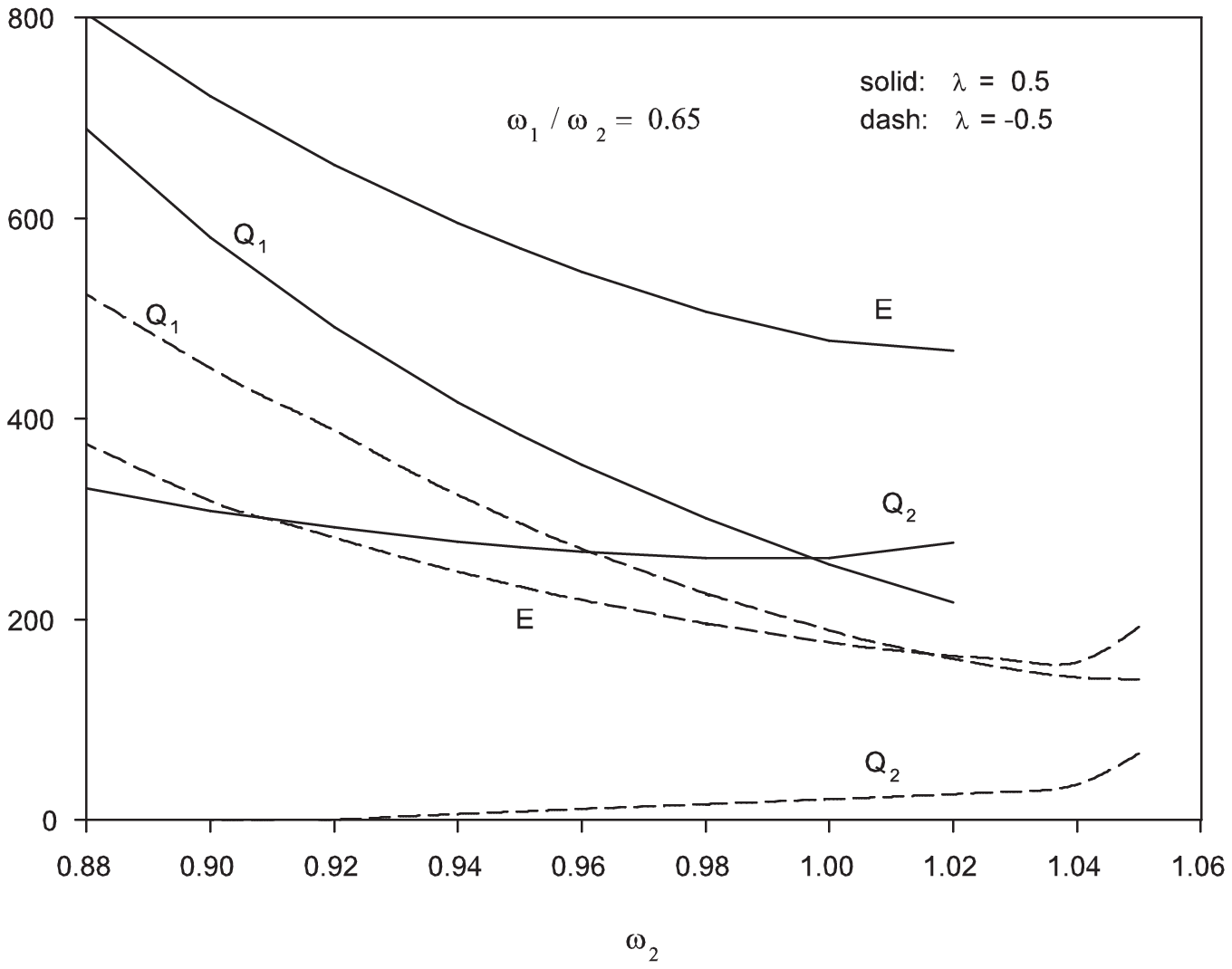}\\
\caption{\label{eqq_mix} The energy $E$ and charges $Q_1$, $Q_2$ are shown
as functions of $\omega_2$ for $\omega_1 = 0.65 \omega_2$ and $\lambda=\pm 0.5$.}
\end{figure}

We have also studied the dependence of the solution's conserved quantities
on $\omega_2 = \omega_1/0.65$ for $\lambda = \pm 0.5$. The dependence of
the energy $E$ and the charges $Q_1$, $Q_2$ is shown in Fig.\ref{eqq_mix}.
These results strongly suggest that for $\lambda<0$ and in the region of the parameter space chosen, the
field $\phi_2$ corresponding to the $k_2=1$ $Q$-ball  tends uniformly to zero for a critical
value of $\omega_2=\omega_2^{(cr)}$ such that
$Q_2\rightarrow 0$ for $\omega_2\rightarrow \omega_2^{(cr)}$.
Only the field $\phi_1$ remains non-trivial when $\omega_2\leq \omega_2^{(cr)}$.
This effect can also be observed in Fig.\ref{t00bis}, 
where the solution for $\lambda=-0.5$ has nearly
lost all its axially symmetric character.

We observe the inverse phenomenon for $\omega_1=c \omega_2$ with a constant $c > 0$.
We don't present our detailed results here since they are qualitatively equivalent to
the case discussed above. We find that $Q_1\rightarrow 0$ for $\omega_1\rightarrow \omega_1^{(cr)}$. Thus, the spherically symmetric solution disappears from the system, while
$\phi_2$ remains non-trivial for $\omega_1 \leq \omega_1^{(cr)}$.

Apparently, while in the case $\omega_1=\omega_2$ and $\lambda < 0$, the charge $Q_1$ associated to the spherical $Q$-ball tends to zero for $\omega_1\rightarrow \omega_1^{(cr)}$,
it is the charge $Q_i$, $i=1,2$ of the $Q$-ball with the higher frequency that tends to
zero for $\omega_i\rightarrow \omega_i^{(cr)}$, $i=1,2$ when $\omega_1\neq \omega_2$ and $\lambda < 0$. Note that nothing similar is observed when $\lambda \ge 0$.

\begin{figure}[!htb]
\centering
\leavevmode\epsfxsize=12.0cm
\epsfbox{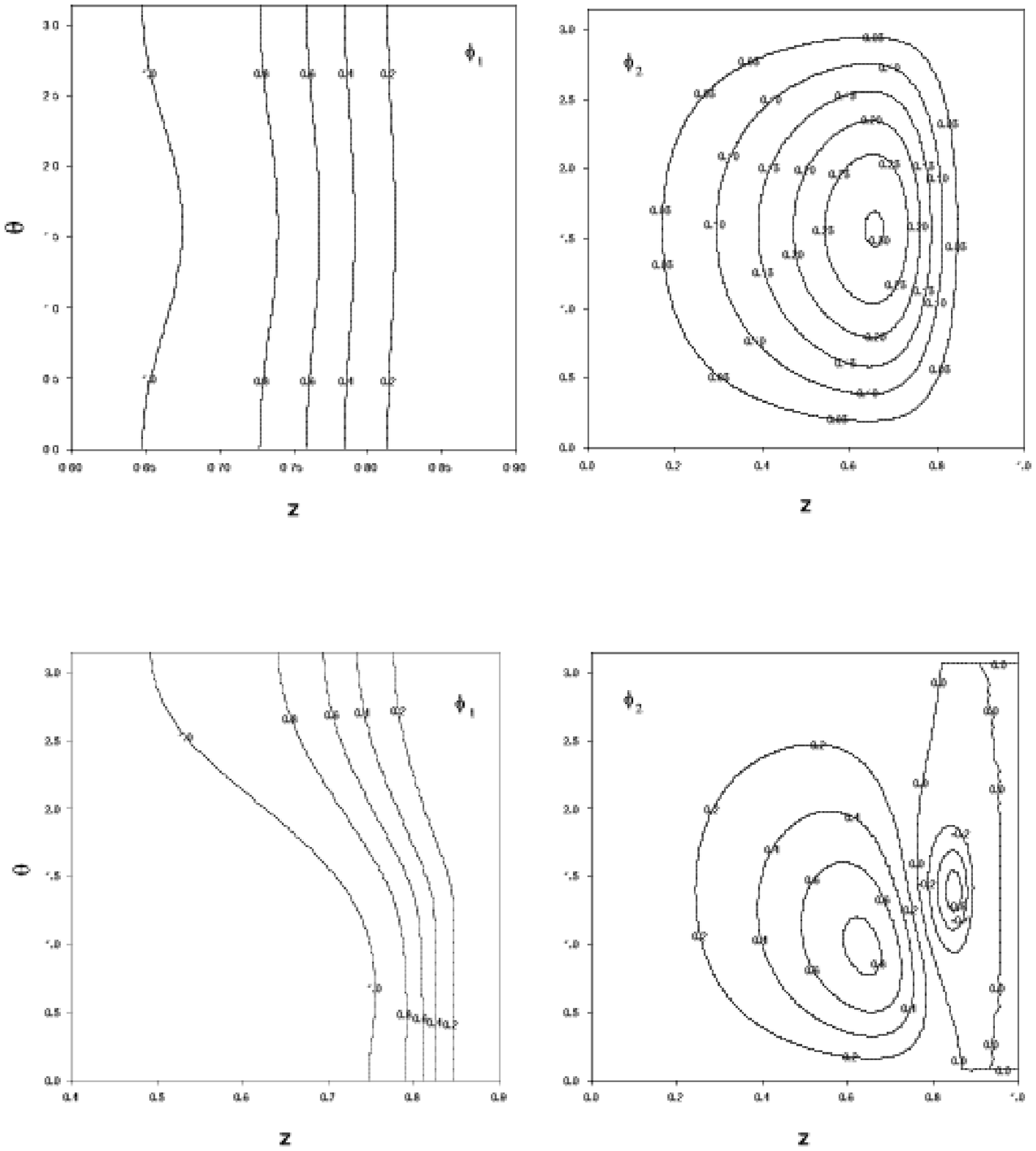}\\
\caption{\label{asym} The contour plots for $\phi_1$ and $\phi_2$ of a parity-even
2-$Q$-ball solution (upper part) and of an asymmetric 2-$Q$-ball solution (lower part) are shown   
for $\lambda = -0.5$, $\omega_1=0.585$ and $\omega_2=0.9$}
\end{figure}

While 1-$Q$-ball solutions known so far are always either parity-even or parity-odd with respect to
$\theta\rightarrow \pi-\theta$, we have constructed several examples of 2-$Q$-ball solutions 
 that do not have a
defined parity. One such solution is shown in Fig.\ref{asym} (lower part) together with
a parity-even solution (upper part). These solutions exist for exactly the same values of the coupling constants.
 Both functions $\phi_1, \phi_2$ are clearly
neither parity-even nor parity-odd and the field $\phi_2$ possesses in addition nodes in the radial direction.
This solution is thus an asymmetric, radially excited 2-$Q$-ball solution. As expected,
we observe that this asymmetric solution has much higher energy and
charges than the corresponding parity-even solution. 

The investigation of 
solutions of this type and their eventual bifurcation into branches of solutions with defined
parity is currently underway.

\section{Concluding remarks}
In this paper, we have presented numerical evidence that $Q$-balls solutions admit
several types of excitations labelled by integers. So far, it was known that the
static, spherically symmetric solution is the ``ground state'' of a series of radially excited solutions.
Families of spinning solutions are also known, they are axially symmetric and can be labelled according 
to the winding $k$ around the axis of symmetry. Here we present evidence that excitation
with respect to $\theta$ can be constructed as well. Generally, the previous results and the present analysis
suggest that families of elementary solutions of the field equations exist and are labelled by $n$, $L$, $k$, where
$n$ refers to the number of nodes in radial direction, while $L$, $k$ refer to the ``quantum numbers'' related to the spherical harmonics.
At the moment, the only analytic argument we have for this property
is its analogy to the linearized version (i.e. small field limit) of the equation  
where this result holds true by standard harmonic analysis. It is likely that the qualitative
properties of the solutions exist also in the case of the full non-linear equations.

We have also studied a system of two interacting $Q$-balls and have constructed
several examples of axially symmetric, stationary solutions that carry conserved currents
and charges. We observe that the 2-$Q$-ball solutions exist in a finite range of the
frequency $\omega_{i,min}\le \omega_i \le \omega_{i,max}$, $i=1,2$, where $\omega_{i,max}$
is independent of the interaction coupling, while $\omega_{i,min}$ does 
dependent of the interaction coupling in a highly non-trivial manner. We find that
the charges $Q_i$, $i=1,2$ of the 2 $Q$-balls in interaction tend to infinity when $\omega_i\rightarrow \omega_{i,max}$ or  $\omega_i\rightarrow \omega_{i,min}$ as long
as $\lambda \ge 0$. For $\lambda < 0$, however, we observe that the charges $Q_i$ 
associated to the $Q$-ball with the higher frequency $\omega_i$ tends to zero for
$\omega_i\rightarrow \omega_{i}^{(cr)} < \omega_{i,max}$. For $\omega_{i,min} \le \omega_i \le \omega_{i}^{(cr)}$ only the remaining field $\phi_j$, $j\neq i$ is non-zero.

In a future publication, we intend to construct solutions with the more realistic potential available from
supersymmetry \cite{kusenko} and put the emphasis on the possibility of constructing $Q$-balls and their
excited and/or spinning versions with potentials involving only quartic terms in the scalar fields.
\\
\\
\\

{\bf Acknowledgments} We thank Y. Verbin for discussions at the first stages of this paper.
Y.B. thanks the Belgian FNRS for financial support.

\end{document}